# Wafer-size $VO_2$ film prepared by water-vapor oxidant

Hui Ren[1,2+], Bowen Li[1], Xiaoyu Zhou[1], Shi Chen[1], Yamin Li[1], Changlong Hu[1], Jie Tian[3], Guobin Zhang[1], Yang Pan[1], Chongwen Zou[1]*

[1]National Synchrotron Radiation Laboratory, University of Science and Technology of China, Hefei, 230029, China

[2]Institute of Scientific and Industrial Research, Osaka University, Ibaraki, Osaka 567-0047, Japan

[3]Engineering and Materials Science Experiment Center, University of Science and Technology of China, Hefei, 230026, P. R. China

*Corresponding Author: czou@ustc.edu.cn

## Abstract

The growth of wafer-scale and uniform monoclinic $VO_2$ film with excellent phase transition property was a challenge if considering the multivalent nature of vanadium atom and the various phase structures of $VO_2$ compound. Direct oxidation of metallic vanadium film by annealing in oxygen gas or in air was suggested to be an easy way for $VO_2$ film preparation, while the oxidation parameters including the gas pressure, gas flux and annealing temperature were extremely sensitive due to the critical preparation window. Here we proposed a facile water-vapor assisted thermal oxidation to produce wafer-scale $VO_2$ films with good uniformity. Results indicated that by using the water vapor as the oxidizing agent, the temperature window for $VO_2$ growth was greatly broadened. In addition, the obtained wafer-size $VO_2$ film showed very uniform surface and sharp resistance change, which was attributed to the improved crystallinity and the high oxygen stoichiometry. The chemical reaction route was calculated based on Gibbs free energy, which confirmed the preferred growth of $VO_2$ film by wet oxidation. Our results not only demonstrated that the water vapor could be used as a modest oxidizing agent, but also showed the unique advantage for large size $VO_2$ film preparation, which was useful for practical devices applications in the future.

# 1. Introduction

As a prototypical transition oxide, vanadium dioxide ($VO_2$) shows a pronounced metal-insulator transition (MIT) behavior at the critical temperature of about 68℃, accompanying by a sharp resistance variation up to five orders of magnitude [1-3] and a distinct infrared switching effect within sub-picosecond time range [4]. These considerable MIT behaviors make $VO_2$ material many potential applications in smart windows [5-7], optical switching devices [8, 9], infrared laser protection [10, 11] and memory devices [12-14].

Since the performance of $VO_2$ based device is directly associated with its MIT behavior, $VO_2$ thin film growth with pure phase has attracted much interest for many years. The quality of $VO_2$ film is very sensitive to oxygen stoichiometry due to the multivalent states of Vanadium ions, which is easy to cause the Magnéli phases such as $V_nO_{2n-1}$ ($3\leq n\leq 9$) [15, 16] and seriously degrades the MIT property. Many methods have been adopted to grow epitaxial $VO_2$ film, such as Molecular Beam Epitaxy (MBE) [17, 18], Pulsed Laser Deposition (PLD) [19-21]. For the preparation of wafer-scale and monoclinic $VO_2$ film with excellent MIT behavior, it is still a challenge if considering the pure phase structure and uniformity. Although it was claimed that MBE technology was able to grow wafer-size $VO_2$ film [3], the process was complex and quite expensive, which was not suitable for quantity device fabrication.

Thermal oxidation of metallic vanadium film in oxygen gas or in air was considered to be a cheap way for $VO_2$ film preparation, while the oxidation parameters, such as the gas pressure, annealing temperature and time, were extremely sensitive due to the critical growth window [22-25]. C.G. Granqvist group reported the $VO_2$ films by thermal oxidation of vanadium in $SO_2$ gas since they considered that the oxidation agent was milder than pure oxygen. The toxic gas used in the experiment and much higher temperature required for the annealing treatment (600~650℃) were obvious disadvantages [26]. Thermal oxidation approaches by water vapor was reported to be effective for metal oxidation treatment [27-30] due to the oxygen partial pressure modulation ability within a controllable temperature range [27, 31-37]. However till now, there was no report about the water-vapor-assisted oxidation for large-size $VO_2$ film preparation directly from metallic vanadium film.

In this work, we achieved a facile water-vapor assisted thermal oxidation of metallic vanadium film at ambient pressure and obtained a homogeneous single phase structure $VO_2$ film with excellent MIT properties. The phase transition behaviors of the obtained $VO_2$ films with or without water vapor were systematically investigated. Results indicated that the water-vapor played very important roles for the thermal oxidation of metallic V film to the final pure $VO_2$ film, which not only broaden the growth window, but also enhanced the uniformity and MIT performance for the obtained large-size $VO_2$ thin film, demonstrating the unique advantage of this wet-oxidation method for scalable $VO_2$ film preparation.

## 2. Experimental Section

Metallic vanadium thin film could be easily deposited by thermal evaporation or magnetron sputter. In the current experiment, we prepared the precursor metallic V thin film on 2-inch c-cut sapphire wafers by thermal evaporation technique. The growth rate was calibrated by a quartz based thickness monitor (Inficon SQM-160) and the final film thickness was kept at ~30nm and 200nm respectively.

To conduct the water-vapor effects on the oxidation of metallic vanadium film, a home-made water-vapor assisted furnace system was established as shown in Figure 1. Argon gas as the carrier gas was controlled by the mass flow controller (MFC) and piped into a half-filled water flask to produce bubbles. The water flask was kept in a water bath, which was heated to produce the water vapor. The amount of water vapor content transported into the furnace was modulated by both the water bath temperature and argon flux [27].The as-fabricated metallic films were oxidized in the above furnace system at different temperatures ranging from 450 to 650ºC by introducing a flow of 0.1~0.8 liter per minute (LBM) argon gas with and without water vapor, respectively. During the wet-oxidation, the water flask half-filled with deionized water was kept in the water bath with the temperature of 50~70ºC. After oxidized in the water vapor for about two hours, the sample was cooling naturally to 150ºC before taken out of the furnace.

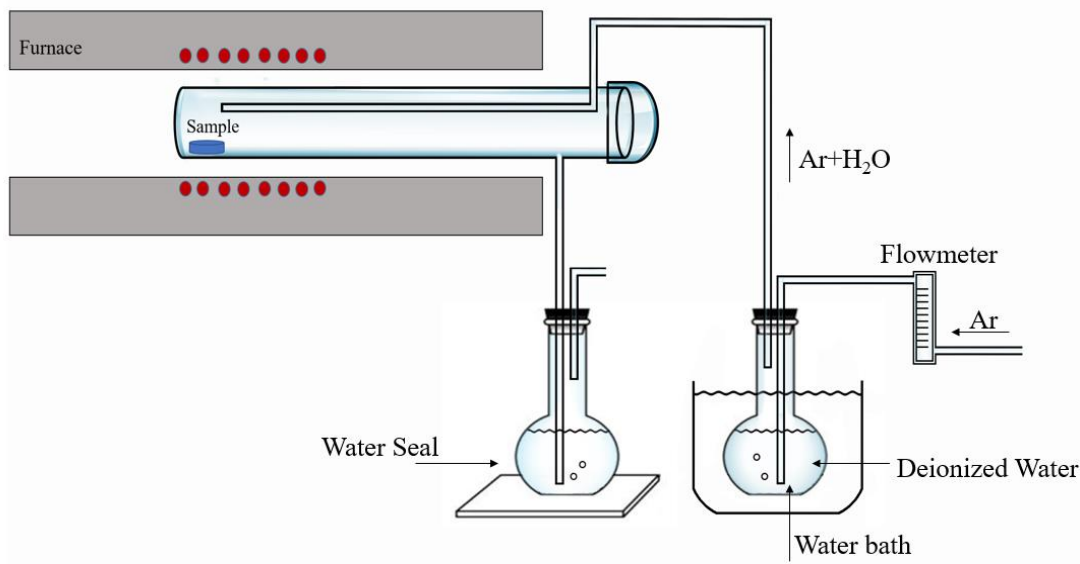

Figure 1. The scheme for the water-vapor assisted oxidation furnace system. Argon gas flux controlled by the Mass Flow Controller (MFC) was used as the carrier gas. The water flask half-filled with deionized water was kept in the water bath for water vapor production.

The prepared samples were characterized by X-ray diffraction (XRD, Model D/Max 2550 V, Rigaku, Japan) with Cu Kα radiation ($\lambda = 1.54178$Å). The θ-2θ scan mode was used with the goniometer precision of about 0.002º. Raman spectra were collected at room temperature by a Raman Spectrometer (LABRAM-HR 750 K, $\lambda = 532$nm, resolution: $0.65cm^{-1}$ (585nm)) with a laser power of 0.5 mW. The X-ray spectroscopy (XPS, Thermo ESCALAB 250, Al Kα 1486.7 eV) was used to examine the chemical states of the $VO_2$ film. During the XPS scan, the CAE analyzer mode was selected with the pass energy of 30.0eV. The scan step was set at 0.05eV. The XPS peak fitting was conducted by XPS Peak 4.1 software package with Gaussian-Lorentzian profile. The electron probe micro-analyzer (EPMA, SHIMADZU EPMA-8050G) was used to examine the surface morphology and the atomic concentration. The V element distribution on the surface was examined by the wavelength dispersive spectra mapping (WDS-mapping). The X-ray absorption near-edge spectroscopy (XANES) was conducted at the XMCD beamline (BL12B) in National synchrotron radiation laboratory (NSRL), Hefei.

## 3. Results and Discussion

After 30-nm $VO_2$ samples were prepared by wet oxidation with different temperatures (carrier gas~0.8 LBM, water bath~50 °C ), the crystal structures were characterized by XRD with the θ-2θ scan mode and the results were shown in Fig. 2(a). It was observed that except the diffraction peak from $Al_2O_3$ (0001) substrate, the unique peak around 2θ=39.89° was assigned to the diffraction from monoclinic $VO_2$ (020) (JCPDS card #82-0661). This observation showed the wet-oxidation treatment can produce pure phase $VO_2$ film and the obtained film was grown with the preferred orientation. Fig. 2(b) showed the related Raman spectra for the obtained 30-nm $VO_2$ samples. The strong Raman peaks at 193, 225, 310, 389, and 613 cm$^{-1}$ were observed, confirming the formation of monoclinic $VO_2$ films by the water-vapor assisted oxidation from metallic V film.

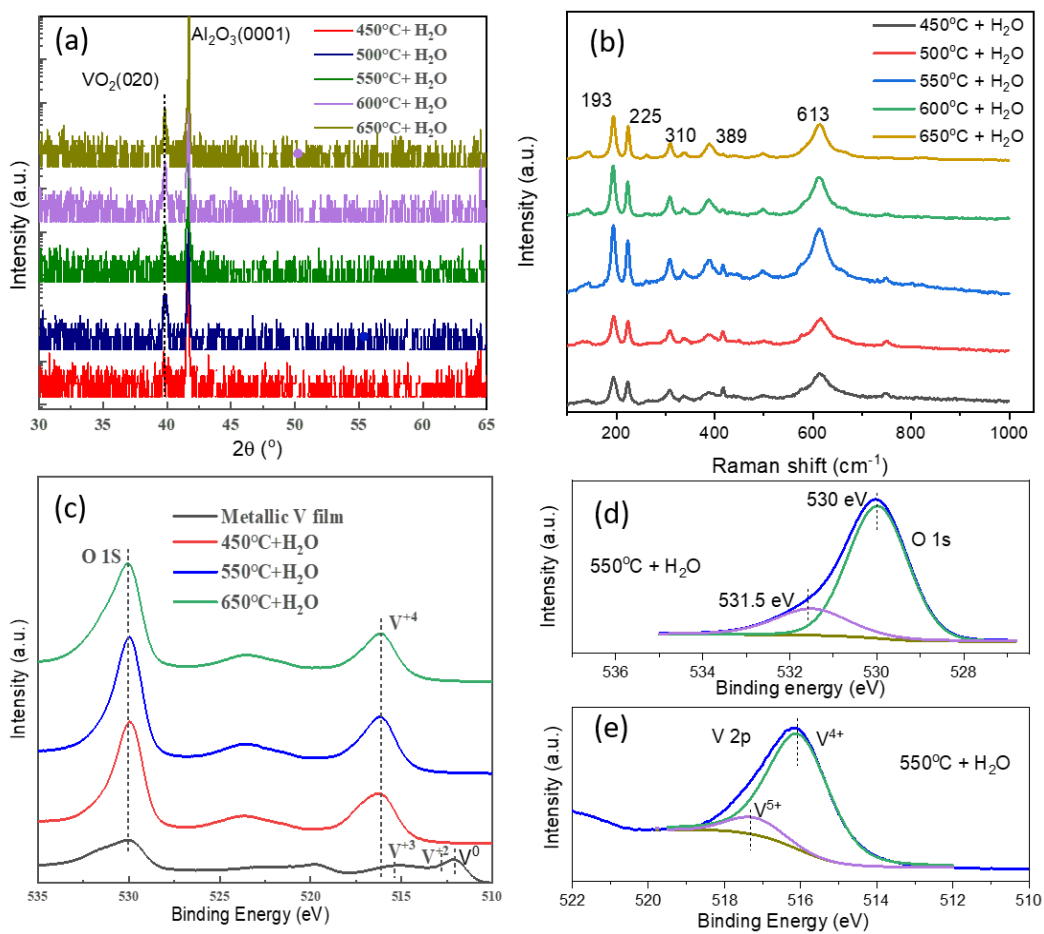

Figure 2. The XRD pattern (a), Raman tests (b) and XPS measurements (c) for the prepared $VO_2$ film by water-vapor assisted oxidation at different temperatures; (d) and (e) showed the curve-fittings of O$1s$ and V$2p_{3/2}$ peaks for the $VO_2$ film at 550 °C.

The chemical state of the obtained $VO_2$ films were examined by X-ray photoelectron spectroscopy (XPS) in Fig. 2(c). From the V2p$_{3/2}$ and O1s peaks for the $VO_2$ samples by wet-oxidation, it was confirmed that the metallic V precursor films were successfully oxidized by water vapor. According to the pronounced V2p$_{3/2}$ peak profiles for the samples, the $V^{4+}$ state remained the major contribution. For the obtained $VO_2$ film at 550°C, the detailed XPS curve-fittings were shown in Fig. 2(d) and 2(e). The O1s peak located at 530eV was assigned to the O-V bonds, while the 531.5eV peak should be from the surface absorbed –OH species. The V2p$_{3/2}$ peak was dominated by $V^{4+}$ state at 516.1 eV [38, 39]，which indicated the pure $VO_2$ film formed. The peak with higher binding energy (~517.2eV) was related to the $V^{5+}$ state, which should be from the surface oxidation when the sample was exposing in air. The dominated $V^{4+}$ chemical states of the obtained $VO_2$ film by wet-oxidation were also verified by the synchrotron based absorption near-edge structure (XANES) spectra (see Supplementary Figure S1). From the above characterizations, it was clear that the obtained film produced by water-vapor at 550°C showed the pure monoclinic phase structure and excellent stoichiometry.

The surface morphology of the $VO_2$ sample by wet-oxidation was also examined by SEM as shown in Fig.3. For the metallic V film, a smooth surface with very small grains was shown, while after wet-oxidized at 450°C, the produced $VO_2$ film had enlarged particle size on the surface. Further increasing the oxidation temperature to 550°C, the $VO_2$ film was still continuous and the surface particles grown much larger. However, if the temperature was further increased up to 650°C, the obtained film was broken with many holes on the surface, which indicated the film was discontinuous. Thus from the SEM observation, it was suggested that wet-oxidation at 550°C should be the suitable condition for $VO_2$ film preparation.

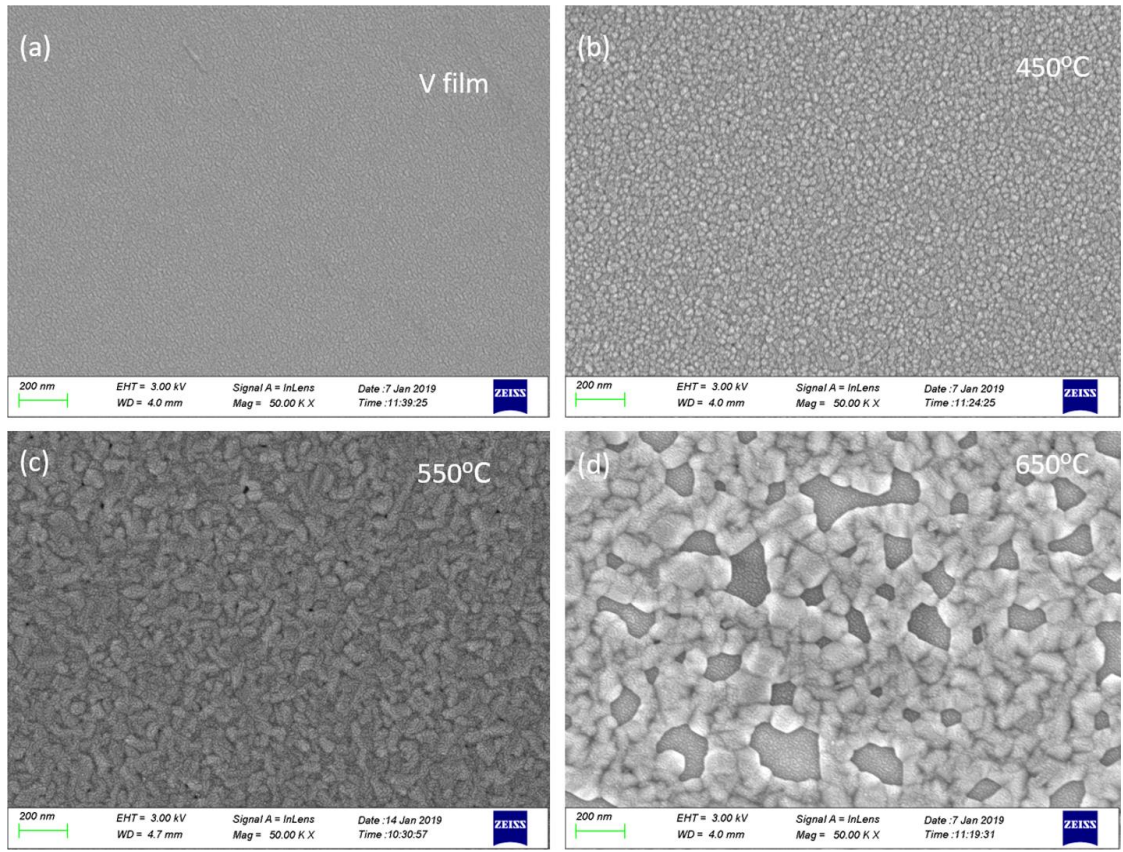

Figure 3. (a) The SEM images for the metallic V film; (b~d) The SEM images for the $VO_2$ film prepared by wet-oxidation at the temperatures of 450ºC, 550ºC, 650ºC, respectively.

The resistance measurements as the function of temperature (or R-T curve) for the obtained 30-nm$VO_2$ films by wet oxidation at different temperatures were also conducted as shown in Fig. 4(a). It can be observed that as the temperature increased from 450ºC to 550ºC, the R-T curve became sharper and the resistance ratio between the insulating metallic phases across the MIT transition was improved. However, if further increasing the wet oxidation temperature up to 600ºC, the resistance change in the R-T curve was decreased. Accordingly, the variations of R-T curves further confirmed that 550ºC should be the optimized wet-oxidation temperature. Fig. 4(b) showed R-T curves for the prepared $VO_2$ film by wet-oxidation and by conventional oxidation treatment in air. Both of the samples were obtained under the optimized conditions. It was evident that the wet-oxidation induced $VO_2$ film demonstrated much better quality according to the sharp MIT curve and the great resistance change up to four orders of magnitude.

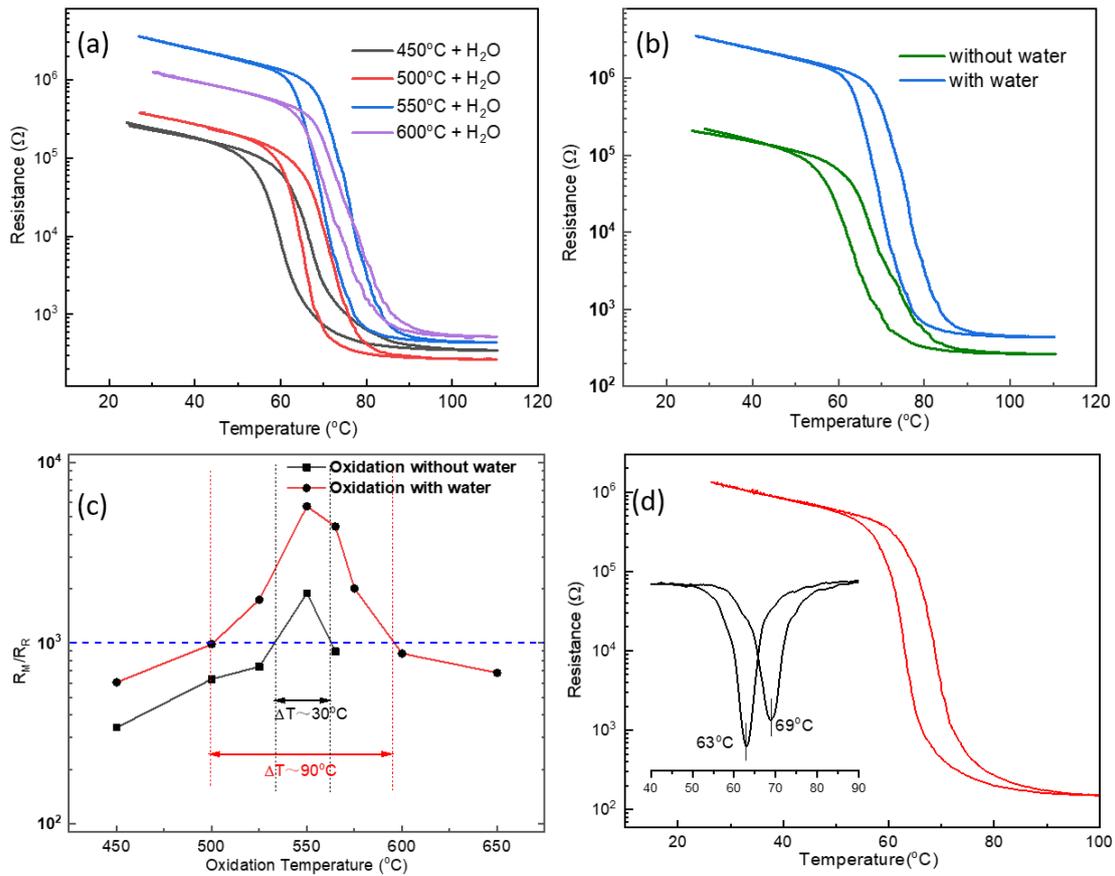

Figure 4. (a) The resistance measurements as the function of temperature (R-T curve) for 30nmVO$_2$ film by wet-oxidation; (b) The R-T curves for the optimized VO$_2$ films by wet-oxidation and by normal oxidation treatment without water-vapor; (c) The resistance ratio between the insulating monoclinic VO$_2$ film and rutile phase ($R_M/R_R$) as the function of oxidation temperature. The wet-oxidation induced VO$_2$ films showed the $R_M/R_R$ value up to $10^3$ with the broadened oxidation temperature window(ΔT from 30°C to ~90°C) . (d) By wet-oxidation route, 200nm VO$_2$ film showed the $R_M/R_R$ value up to four order of magnitudes. The insert showed the related differential curves.

In addition, we also found that the annealing temperature window was greatly broadened to produce pure VO$_2$ film by wet-oxidation treatment as shown in Fig.4 (c), in comparison with the conventional oxidation method in air or in oxygen gas. For conventional oxidation, VO$_2$ film could be only obtained by very harsh conditions and the annealing temperature window (ΔT) was only ~30°C if keeping the resistance change to three orders of magnitude across the phase transition as reference value. Interestingly, by wet-oxidation in water vapor, the oxidation temperature window was broadened up to 90°C for $R_M/R_R$ value >$10^3$, which made it easier to control the growth conditions. Furthermore, by wet-oxidation method, we could produce thick $VO_2/Al_2O_3$ film with clear cross-section (see Supplementary Figure S2). In Fig.4 (d), the obtained 200nm $VO_2/Al_2O_3$ film showed quite sharp MIT properties and the resistance change was up to four orders of magnitude. From the related differential curves as shown in the insert, the critical temperatures were determined to be 69°C for the heating process and 63°C for the cooling process, respectively, which were quite consistent with previous reports [27, 40, 41] by other deposition techniques.

Wet-oxidation route can also be used to prepare uniform $VO_2$ film at large-size. Fig.5 (a) showed the prepared 2-inch metallic V precursor film. After treated by wet-oxidation at 550°C, pure $VO_2$ film can be obtained as shown in Fig.5 (b). A 2-inch $VO_2$ film was produced with good uniformity from the eye-sight view. To further examine the uniformity of the obtained $VO_2$ film, we measured the surface distribution of resistance values before and after the MIT process (at 25°C and 90°C respectively), and the results were plotted in Fig. 5(c) and 5(d), respectively. It was observed that the insulating monoclinic $VO_2$ film showed the resistance values of about 0.5~1.7 MΩ within the 2-inch surface at room temperature, while the resistance was sharply decreased to 100~140Ω at 90°C over the whole surface. The resistance change during the MIT process was larger than three orders of magnitude for the whole 2-inch sample, indicating the excellent uniformity of the obtained $VO_2/Al_2O_3$ film sample. However, by conventional oxidation method in oxygen or in air, the obtained 2-inch $VO_2$ film was always inhomogeneous (Supplementary Figure S3), since its uniformity was poor even optimizing the oxidation condition.

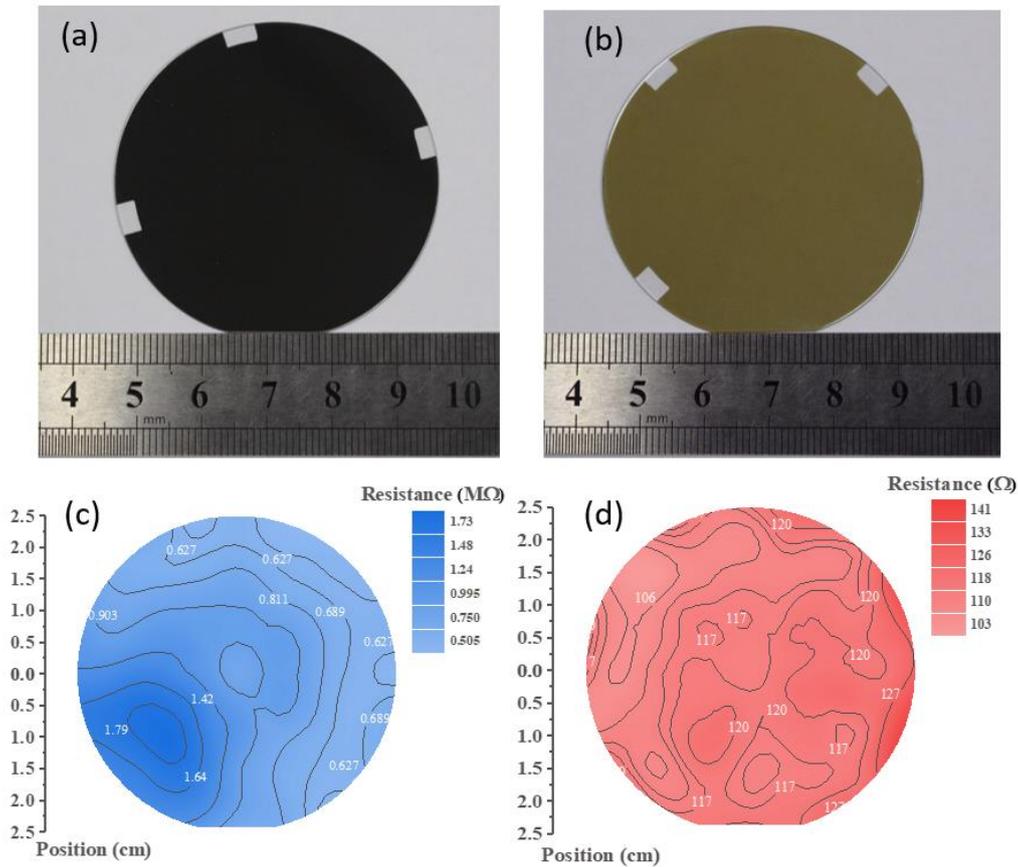

Figure 5. (a) The metallic V film deposited on sapphire substrate with 2-inch wafer-size; (b) After oxidized by water-vapor at 550°C, the uniform 2-inch $VO_2$ film was produced; (c) and (d) The surface resistance distribution for the obtained $VO_2$ film before (25°C) and after (90°C) the phase transition, confirming the uniformity of the obtained 2-inch $VO_2$ film.

In addition, the wet-oxidation by water vapor is also suitable for large-size $VO_2$ film on other substrates. It is known that there exists large lattice mismatching between VO$_2$ and Si crystal, thus epitaxial VO$_2$ film growth on Si substrate is always difficult to obtain. By our wet-oxidation method, 200 nm thick VO$_2$ /Si(001) film with 2-inch size was prepared as shown in Fig.6 (a). The prepared wafer-size VO$_2$ film was quite uniform even by eyesight. The quite sharp R-T curve in Fig.6 (b) and the distinct Raman spectra in Fig.6 (c) further confirmed the good quality and uniformity of the prepared VO$_2$ /Si(001) film by water-vapor oxidation. While the conventional oxidation method induced VO$_2$ /Si(001) film in oxygen gas or in air with optimized conditions was always inhomogeneous( Supplementary Figure S4 and Table 1S). Furthermore, large size and uniform VO$_2$ /mica film with excellent flexibility was also prepared by our water-vapor oxidation method as shown in Fig.6 (d), demonstrating the universal of such wet-oxidation technique for VO$_2$ film preparation.

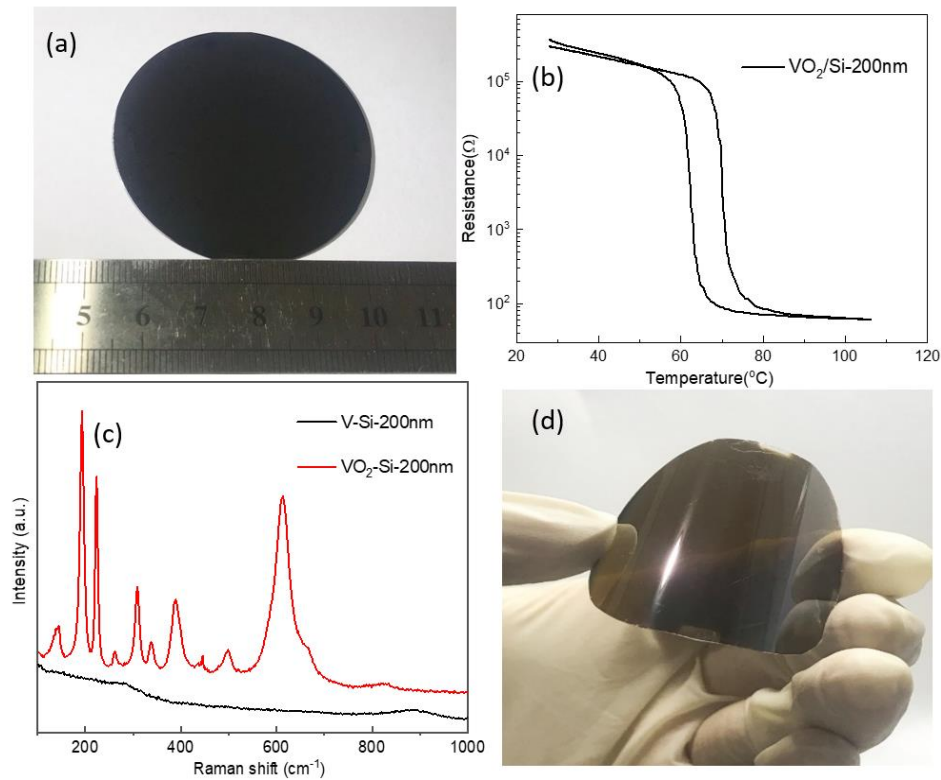

Figure 6. (a) The prepared 2-inch 200nm $VO_2$ /Si(001) film by water-vapor oxidation; (b) and (c) The R-T curve and Raman spectra for the $VO_2$ /Si(001) sample; (d) The flexible and uniform $VO_2$ /mica film with 2-inch by water-vapor oxidation

From the experimental results, it was inferred that the wet-oxidation, or the water vapor assisted oxidation process showed various advantages, comparing with conventional oxidation way (Supplementary Figure S5). In general, V atoms showed multi-valance states from 0, +1, +2…+5 valance states, thus it was very sensitive to the oxidation conditions, which determined the final vanadium oxides. While it was suggested that water-vapor could be acted as a modest oxidizing agent, which was not active enough to oxidize V$^{4+}$ ions to V$^{5+}$ states, thus there existed an ideal self-adjustment of oxygen pressure when annealing the metallic V precursor in water-vapor.

In fact, the reaction between water vapor and metallic V film can be written as: 2H$_2$O (g) + V = 2H$_2$ (g) + VO$_2$. While the traditional oxidation route by oxygen gas is: O$_2$ +V=VO$_2$. Accordingly, from the view point of thermo-dynamics, these reactions can be estimated by the system Gibbs free energy: $\Delta G = \Delta H - T \Delta S$. By using the general software HSC CHEMISTRY 6.0, we have calculated the Gibbs free energy for the reactions between water vapor/oxygen gas and metallic V film (Supplementary Table 2S). From the $\Delta G$ values for the final products of VO$_2$ and V$_2$O$_5$ with the temperature of 550$^\circ$C as summarized in Table 1, it was deduced that for the reaction between water vapor and metallic V at 550$^\circ$C, the final product would prefer to form VO$_2$, while not V$_2$O$_5$ compound, since the $\Delta G$ value to form VO$_2$ was -39.9 kcal, which was quite smaller than that of $V_2O_5$ ($\Delta G$=-21.4 kcal). However, for the traditional oxidation route by oxygen gas, the $\Delta G$ values for $VO_2$ and $V_2O_5$ formation were quite close, or even the $\Delta G$ value for $V_2O_5$ was smaller, which indicated that the reaction preferred to $V_2O_5$ formation. Since the final product by the oxidation of metallic V film had no obvious selection to $VO_2$ compound, it was difficult to prepare pure $VO_2$ by this reaction. These results were quite consistent with our experimental observations that the oxidation temperature window was quite narrow and the obtained 2-inch $VO_2$ film was not uniform by the traditional oxidation route in oxygen gas or in air.

Table 1. The calculated Gibbs free energy for the reactions between water vapor/oxygen gas and metallic V film at 550°C

|  | $\Delta G$ ($VO_2$)@550°C | $\Delta G$ ($V_2O_5$) @550°C |
| --- | --- | --- |
| $H_2O(g)$ +V | -39.9kcal | -21.4kcal |
| $O_2(g)$+V | -136.6 kcal | -142.3 kcal |

In addition, it was reported that the reaction of $H_2O$ molecules with the surface of oxides [27] was more easily to occurrence than oxygen molecules in a dry environment at high temperature, which was usually known as the over-reduced environment. Thus it was deduced that during the wet-oxidation process at high temperature, $H_2O$ molecules was adsorbed and chemically dissociated into $OH^-$ and $H^+$ on the oxides surface, perhaps due to the possible catalyst effect of vanadium oxides [42]. Then the O-H bonds were further broken to form free oxygen ions to serve as the oxidizing agent [43]. While some H atoms diffusion in the $VO_2$ crystal lattice would be favor to remove some crystal defect such as the oxygen vacancies [31], which further improve the quality of $VO_2$ film.

## 4. Conclusions

In summary, we have successfully prepared large-size $VO_2$ film by facile wet-oxidation technique. In this route, the water-vapor was used as a modest oxidizing agent and metallic V precursor film could be effectively oxidized to monoclinic $VO_2$ film with pure phase. Comparing with traditional oxidation method for $VO_2$ film preparation, the proposed wet-oxidation way not only improved the $VO_2$ film quality and uniformity, but also broadened the annealing temperature window greatly, which make the wafer-size $VO_2$ film preparation more easier and effective. It was suggested that the advantages of wet-oxidation method were mainly lying on the facile oxidation ability of water as well as the easy facility. The current studies supply a simple $VO_2$ film preparation with large-size, which should be meaningful for further $VO_2$ based devices applications in the future. Furthermore, this simple and reproducible wet-oxidation technique demonstrates that the common water vapor can be used as a modest oxidizing agent, which should be applicable for prepare some other transition metallic oxides with multi-valance states.

**Acknowledgements**

This work was partially supported by the National Key Research and Development Program of China (2016YFA0401004), the National Natural Science Foundation of China (11574279, 11704362), the funding supported by the Youth Innovation Promotion Association CAS, the Major/Innovative Program of Development Foundation of Hefei Center for Physical Science and Technology and the China Postdoctoral Science Foundation (2017M622002). This work was partially carried out at the USTC Center for Micro and Nanoscale Research and Fabrication. The authors also acknowledged the supports from the Anhui Laboratory of Advanced Photon Science and Technology. The approved beamtime on the XMCD beamline (BL12B) in National Synchrotron Radiation Laboratory (NSRL) of Hefei was also appreciated.

**Competing Interests**

The authors declare that they have no competing financial interests.